\documentclass[journal,12pt,draftclsnofoot,onecolumn]{IEEEtran}
\usepackage{times,epsfig,epstopdf,latexsym,multirow,array,amssymb,graphicx,multicol,stfloats,textcomp,psfrag}
\usepackage[cmex10]{amsmath}
\usepackage{multirow}
\usepackage[english]{babel}
\usepackage{algorithm}
\usepackage{algorithmic}
\usepackage{amsfonts}
\usepackage{bm}
\usepackage{here}
\usepackage{rawfonts}
\usepackage[latin1]{inputenc}
\usepackage[T1]{fontenc}
\usepackage{calc}
\usepackage{url}
\usepackage{enumerate}
\usepackage{color}
\usepackage{upref}
\usepackage{epic,eepic}
\usepackage{dsfont}
\usepackage{comment}
\usepackage{cite}

\DeclareMathOperator*{\argmax}{arg\,max}

\newtheorem{theorem}{Theorem}

\newtheorem{lemma}{{Lemma}}
\newtheorem{corollary}{{ Corollary}}

\newtheorem{remark}{{\bf Remark}}
\usepackage{float}
\usepackage{stfloats}

\DeclareMathOperator*{\argmin}{\arg\!\min}

\begin{document}

\title{\LARGE{Performance Analysis for Energy Harvesting Communication Protocols with Fixed Rate Transmission} }

\author{Mahmood~Mohassel~Feghhi,
        Aliazam~Abbasfar,~\IEEEmembership{Senior~Member,~IEEE},
        and~Mahtab~Mirmohseni
\thanks{M. Mohassel Feghhi and A. Abbasfar are with the School of Electrical and Computer Eng., College of Eng., University of Tehran, Tehran 14395-515, IRAN (e-mail: {mohasselfeghhi, abbasfar}@ut.ac.ir).}
\thanks{M. Mirmohseni is with the Department of Electrical Engineering, Sharif University of Technology, Tehran, IRAN (e-mail: mirmohseni@sharif.edu).}}
%
%
\maketitle

\begin{abstract}
Energy Harvesting (EH) has emerged as a promising technique for Green Communications and it is a novel technique to prolong the lifetime of the wireless networks with replenishable nodes. In this paper, we consider the energy shortage analysis of fixed rate transmission in communication systems with energy harvesting nodes. First, we study the finite-horizon transmission and provide the general formula for the energy shortage probability. We also give some examples as benchmarks. Then, we continue to derive a closed-form expression for infinite-horizon transmission, which is a lower bound for the energy shortage probability of any finite-horizon transmission. These results are proposed for both Additive White Gaussian Noise (AWGN) and fading channels.
Moreover, we show that even under \emph{random energy arrival}, one can transmit at a fixed rate equal to capacity in the AWGN channels with negligible aggregate shortage time. We achieve this result using our practical transmission schemes, proposed for finite-horizon.
Also, comprehensive numerical simulations are performed in AWGN and fading channels with no Channel State Information (CSI) available at the transmitter, which corroborate our theoretical findings. 
Furthermore, we improve the performance of our transmission schemes in the fading channel with no CSI at the transmitter by optimizing the transmission initiation threshold.
\end{abstract}
\section{Introduction}
\IEEEPARstart{E}{nergy} Harvesting (EH) has emerged as a promising solution to the perennial energy constraint of wireless networks, which have low cost mobile devices equipped with fixed energy supplies such as limited batteries \cite{sudev}. The main drawback of such devices is that replacing their batteries is either expensive or impossible, especially in harsh environments or battlefields. Also, EH is developed to be used as a foundation of green communication networks in order to reduce the increasing energy consumption of human activities \cite{zhu}. Energy harvesters collect ambient energy from the environment or harness energy from the human body and convert it into usable electric power. Conventional energy harvesters are solar cells, piezoelectric cells, microbial fuel cells, thermo-electric generators, wind turbines, water mills, vibration absorption devices, etc \cite{sudev}. Although EH nodes have access to an inexhaustible energy source in contrast to conventional battery-powered nodes, limited EH production rate necessitates sophisticated utilization of the scavenged energy for reliable and stable operation of such nodes.

\subsection{Related Work and Motivation} 
Recently, EH green communication networks has gained tremendous attention. Early works on EH sensor nodes are presented in \cite{raghun, kansal, sharma}. In \cite{rajesh}, Shannon capacity of EH sensor nodes in an Additive White Gaussian Noise (AWGN) channel is derived, while inefficiencies in energy storage are studied. 
The authors in \cite{yang} consider a wireless single user EH communication system, in which energy and data packets are stochastically arrived at the source node. They provided the optimal packet scheduling by changing the rate adaptively according to data and energy traffics.
 The study in \cite{yang} is then extended to multiuser channels, such as multiple-access channel \cite{yang_jcn}, broadcast channel \cite{yang_ozel, antepli}, interference channel \cite{tutun}, and a two-hop channel \cite{gunduz}. In \cite{ozel_Jsac}, optimization of point-to-point data transmission with finite battery EH nodes in a wireless fading channel is investigated; directional water-filling algorithm is introduced for optimal offline solution and stochastic dynamic programming is used for optimal online solution. The save-then-transmit EH protocol in slotted transmission mode is studied in \cite{Luo2013},\cite{wcnc2013}: in each slot, a fraction of time (called the \emph{save-ratio}) is solely devoted to energy harvesting while the remaining fraction is used for information transmission. To find the optimal save-ratio, the outage probability is minimized and the throughput is maximized in \cite{Luo2013} and \cite{wcnc2013}, respectively.  
Another line of research focuses on collecting energy from the ambient Radio Frequency (RF) signals and considers the problem of simultaneous wireless information and power transfer over the wireless channels (see e.g. \cite{zhang-WIPT} and references therein).

In recent years, the demand for higher data traffic on mobile operators has significantly increased and it is anticipated to grow further \cite{cisco}, \cite{Fehske}. 
This would motivate to explore novel \emph{energy efficient} techniques in the energy constrained scenarios to achieve the highest data rate possible.
Adaptive Modulation and Coding (AMC) can potentially provide high data rates required in such networks. For this purpose, AMC  requires exact knowledge of Channel State Information (CSI) at the Transmitter (Tx) \cite{AMC}. In most wireless systems, CSI is estimated at the receiving nodes and is provided to the transmitting nodes through feedback links. However, rapid channel variations caused by fading and feedback constraints \cite{feedback} together with complexity and robustness issues of channel prediction schemes make the AMC infeasible in many applications, especially in energy constrained scenarios. In such cases, it is favourable to use a fixed modulation along with a fixed-rate coding selected at the start of communication.

Another motivation for using fixed rate transmission is the EH communication systems with Quality of Service (QoS) constraints. This means that each user has a minimum rate constraint ($R_{\mathtt{target}}$). In fact, a rate guarantee may be required by some delay-sensitive applications in contrast to the best effort services that may not impose any rate constraints \cite{Seong-selected, Seong-isit}. This can be modelled by the following optimization problem:
\begin{equation}
\begin{array}{l} 
{\rm{minimize}}\,\,\,\,\,{P_{out}} = \Pr \left\{ {R < {R_{\mathtt{target}}}} \right\} \yesnumber\\
{\rm{subject}}\,\,{\rm{to}}\,\,\,\,C1:\;\;\;\sum\limits_{i = 1}^j {{p_i}{l_i}}  \le \sum\limits_{i = 0}^{j - 1} {{E_i}} ,\,\,\,\,j = 1,...,M \\
\;\;\;\;\;\,\,\,\,\,\,\,\,\,\,\,\,\,\,\,\,\,\,\,\,\,C2:\;\;\;{p_i} \ge 0,\,\,\,\,\,\,\,\,\,\,\,\,\,\,\,\,\,\,\,\,\,\,i = 1,....,M\, \yesnumber\\
\;\;\;\;\;\,\,\,\,\,\,\,\,\,\,\,\,\,\,\,\,\,\,\,\,\,C3:\;\;\;p = g(R),\,\,g(.)\,\,{\rm{is \,\,monotonically\,\, increasing\,\,function}}\,\,\,\yesnumber
\end{array}
\end{equation}
where $E_i$ is the amount of harvested energy at time instant $T_i$; $l_i$ is the length of time slot $i$ ($l_i=T_i-T_{i-1}$), in which power $p_i$ is used for data communication; and $M$ is the number of time slots. $C1$ (the energy causality constraint) states that the energy cannot be utilized before it is scavenged from the environment. 
Since sending at any rate lower than ${R_{\mathtt{target}}}$ leads to outage, it is suboptimal. Also, sending at any rate higher than ${R_{\mathtt{target}}}$ is suboptimal. The reason is that the power is monotonically increasing function of rate ($C3$). 
 Intuitively, it wastes the energy at no benefit in term of outage. Therefore, the optimal solution should use fixed rate transmission at $R=R_{\mathtt{target}}$.

However, the most important limiting issue in using adaptive schemes is their \emph{complexity}. The complexity constraint may strictly enforce the system to use only one modulation and one coding scheme. This simplifies the EH Transmitter (EH-Tx) and decreases the cost of deploying such transmitting nodes; especially for networks of small sensor nodes. This results in a communication system with fixed rate transmitting nodes, which is emphasized to be a practically and theoretically interesting problem in \cite{Jamali2014}.
\vspace{-0.1cm}
\subsection{Contributions}
\vspace{-0.1cm}
In this paper, we consider \emph{fixed rate transmission} in communication systems with EH nodes in AWGN and fading channels. The fixed rate transmission differentiates our system model from the ones in the literatures on EH green networks, where the Tx can use variable rates for transmission. This scenario imposes a new constraint on our power allocation optimization problem. 
We consider two types of schemes by assuming the availability of the EH profile at the Tx: $i)$ in offline scheme, the Tx knows the EH profile non-causally. $ii)$ in online scheme, the Tx has only access to causal knowledge of energy arrivals.
We introduce Energy Shortage Probability (ESP) as a metric to analyse the system performance. ESP determines the probability of not transmitting due to the energy shortage. It should be noted that if the available energy is lower than the amount required to transmit at the determined fixed-rate, nothing will be transmitted. We are interested in deriving the maximum fixed rate, at which the ESP is minimized.

We adopt the achievable rate scheme of \cite{ozel_TIT} and come up with a practical online finite-horizon (finite transmission time) transmission protocol, called \emph{pause-and-transmit}. In this scheme, if the Tx has enough energy, it sends at rate $R$. Otherwise, the transmission is paused till the required energy for transmission at the predefined rate is available at the Tx. This means that we either send at a fixed rate ($R$) or do not transmit any data (transmission at zero rate). 
The \emph{pause-ratio} is the fraction of time at each time duration of interest that the transmission is impossible due to the lack of energy. This parameter is a stochastic variable that varies from one duration to another.
In this setting, effective transmission rate can be expressed as\footnote{In information theory terminology, this can be achieved by simply time-sharing between a fixed rate, R, and a zero rate, as follows:
\vspace{-0.1cm}
\begin{equation*}
\mathcal{R}_{\mathtt{eff}}=\frac{T-T_S}{T}\times R+\frac{T_S}{T}\times 0.
\end{equation*}
\vspace{-0.1cm}                         
} 
\begin{align}
\mathcal{R}_{\mathtt{eff}}=R(1-\frac{T_S}{T}),
\end{align}
where $T$ is the transmission time of the interest and $T_S$ is the aggregate energy shortage time (time period that transmission is paused due to the lack of sufficient energy).
We also adopt save-and-transmit achievable rate scheme of \cite{ozel_TIT} with some technical differences as an offline finite-horizon transmission protocol, called \emph{store-and-transmit}. In this scheme, the Tx stores the scavenged energy for a fraction of total transmission time until it is able to communicate at the fixed predefined rate ($R$) for the remaining fraction of time. This waiting time can be computed using the non-causal knowledge at the Tx.

Our main contributions are summarized as follows:

\begin{itemize}
\item We show that our problem setting is an interesting case, where our finite-horizon offline and online schemes lead to the same ESP in AWGN channels. In other words, having non-causal information about EH pattern does not improve the performance of optimal scheme in AWGN channels. Since investigating the offline scheme is simpler than the online scheme, we study offline scheme and use its results for the practical online scheme.

\item For finite-horizon transmission, in order to study short-term performance behaviour of the system, exact theoretical expressions for ESP and the effective rate in AWGN channels are derived. These expressions are validated by numerical simulations. Similar results are derived for fading channels. Our results include the exact solutions for the cases with one and two energy harvesting instants, as well as an upper bound on the ESP of any finite-horizon transmission. However, finding the optimal rate and its associated ESP in general is mathematically cumbersome and it is an open problem. Therefore, we analyse the asymptotic performance behaviour of the fixed rate EH-Tx by considering \emph{infinite-horizon} transmission.

\item For infinite-horizon transmission, closed-form solutions for the ESP and for the effective rate are presented in AWGN channel. The former makes a lower bound on the ESP and the latter forms an upper bound on the effective rate for any finite-horizon transmission. These bounds are quite tight in a moderate transmission period.

\item In our scenario with stochastic energy arrival, we show that our \emph{pause-and-transmit} scheme asymptotically achieves the capacity of AWGN channel under average power constraint. Note that we use pause-and-transmit scheme at fixed rate, which is a practical and optimal way of data transmission for systems with EH nodes. Thus, we achieve the capacity using a completely different method compared to impractical achievable coding schemes of \cite{ozel_TIT}.

\item In the case of fading channels, we consider the setups with no Channel State Information at the Transmitter (CSIT). We investigate the outage performance of the optimal online scheme and compare it with our theoretical results. In this case, both the energy shortage and the channel fading degrade the performance of the system (the outage probability is chosen as the performance metric). In addition, we propose a novel online scheme by optimizing the channel power gain threshold for initiating the transmission. This optimization is performed using Channel Distribution Information at the Transmitter (CDIT). Our numerical simulations verifies the theoretical results and shows the superiority of this novel online scheme. 

\item Our results hold for any general power consumption model, which is accomplished by assuming an arbitrary monotonic relationship between consumed power and rate. 
\end{itemize}

The remainder of the paper is organized as follows. The system model and problem formulation is described in Section \ref{section-sys model}. Energy shortage analysis of finite-horizon transmission in AWGN channels is presented in Section \ref{section-finite trans}. Asymptotic behaviour of ESP for infinite-horizon transmission in AWGN channels is investigated in Section \ref{section-asymptotic}. Section \ref{section-fading} provides the optimal online scheme and theoretical expressions for outage probability in fading channels. Theoretical results along with performance bounds are evaluated through numerical simulations in section \ref{section-numeric result}, followed by concluding remarks in Section \ref{section-conclusion}.

\textbf{Notations:} ${\left\lceil x \right\rceil ^ + } = \max \{ x,0\} $, $\bar{f}=\mathbb{E}{f}$, $\mathbb{E}(.)$ is the expectation operation, $\Pr(.)$ is the probability function, and $Q(.)$ is the $Q$-function ($Q(x) = \frac{1}{{\sqrt {2\pi } }}\int_x^\infty  {{e^{ - \frac{{{u^2}}}{2}}}du} $).

\section{System Model and Problem Formulation} \label{section-sys model}
\begin{figure}[t!]
  \centering
  \vspace{-.1cm}
  \includegraphics[width=4.5in]{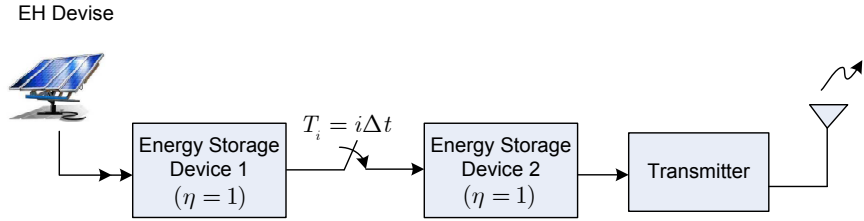}\\
  \vspace{-.1cm}
  \caption{System model.
  \vspace{-0.1cm}
  }\label{system model}
\end{figure}

\vspace{-0.1cm}
We consider a point-to-point communication system, where Tx sends information at a fixed rate and it is capable of being recharged by the ambient energy. The energy harvester device at Tx absorbs stochastic continuous amount of environmental energy and stores it in an Energy Storage Device (ESD). The efficiency of an ESD, $\eta$, is defined as the ratio of the energy released from the ESD to the energy harvested from the environment. Two ESDs with high efficiencies ($\eta=1$), such as super-capacitors \cite{super-capasitor} are employed. This is consistent with the existing models \cite{Luo2013}, which is due to the energy half-duplex constraint. This means that a battery cannot be charged and de-charged simultaneously. The first one (ESD1) is connected directly to the EH device. The stored energy in this ESD is dumped at each $\Delta t$ second ($T_i = i\Delta t,\,\,\,i = 0,1,2,...,M-1$) and transferred to the second ESD (ESD2). The time slot between two consecutive energy arrivals at the ESD2, with the length of $\Delta t$ second, is called an \emph{epoch}. The ESD1 is always in the charging mode and it transfers the charged energy to the ESD2 in negligible time. The ESD2 is connected directly to the Tx; it is always decharging except the negligible time required for transferring the stored energy of ESD1 within an epoch. We consider finite-horizon transmission of length $T=M\Delta t$, e.g., an environmental monitoring sensor reporting the catastrophic conditions of a zone at some finite duration. The Tx uses the harvested energy as its sole source of energy for transmission and hardware power consumption. In our system model, depicted in Fig.~\ref{system model}, energy enters the ESD2 in discrete quantities only at the beginning of each epoch and is used during each epoch. Also, the remaining energy at the end of each epoch is not discarded and can be utilized in later epochs (in contrast to \cite{Luo2013}). We assume that the capacity of the ESDs is much larger than the average EH rate, $B_{max}\gg\bar E$.\footnote{We assume that average EH rate is fixed in the period of transmission ($T$), which is a logical assumption considering our finite-horizon transmission (short-term performance bahavior).}

\begin{figure}[t!]
  \centering
  \vspace{-.1cm}
  \includegraphics[width=3.5in]{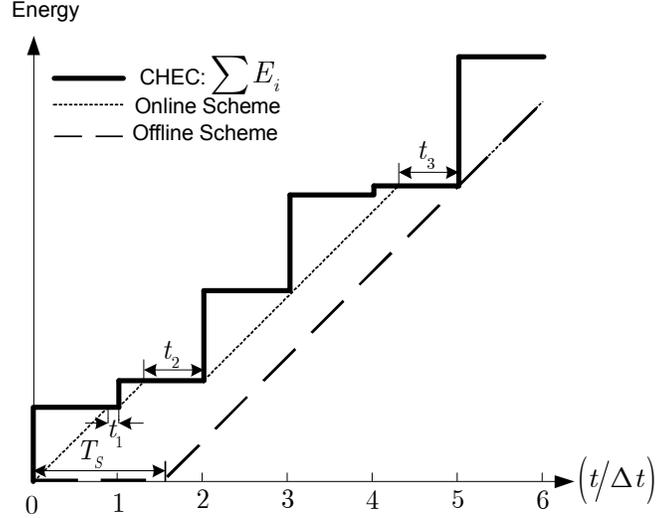}\\
  \vspace{-.1cm}
  \caption{Schematic diagram of energy consumption in online and offline transmission schemes with respect to the cumulative harvested energy curve (CHEC).
  \vspace{-0.1cm}
  }\label{Fig1}
\end{figure}

Our problem is to formulate the probability of energy shortage when the EH-Tx is constrained to send information at a fixed rate. In the online power allocation scheme, the Tx starts to send data at the rate $R$ until its battery is depleted of harvested energy. Then, it waits for the next harvesting instant to resume sending its data to the Receiver (Rx).  
The optimal power allocation solution of our fixed rate transmission protocol with energy causality constraints is not unique. In fact, along with a single optimal online (pause-and-transmit) solution, there are many optimal offline solutions. Consider the energy consumption diagram\footnote{This diagram shows the amount of consumed energy at any time instant.} in Fig.~\ref{Fig1}. We can always find a single non-decreasing offline solution by shifting the harvested energy to the right without violating the energy causality constraint. So we propose an offline scheme that consists of a single store time-slot (to store enough ambient energy) and a single transmit time-slot (to send information at a predefined fixed rate). As can be seen in this figure, both online and offline solutions have the same \emph{aggregate energy shortage time} (the time that Tx does not send due to the lack of energy), i.e., $T_S=t_1+t_2+t_3$. 
It should be noted that in the offline store-and-transmit scheme, the transmission is not necessarily initiated at the beginning of an epoch (see Fig.~\ref{Fig1}).  

The optimal online solutions for the power allocation problem in EH communication systems, requiring only causal knowledge of EH profile, are practically feasible. However, as they typically involves dynamic programming \cite{ozel_Jsac}, their computational complexity is prohibitive even for simple point to point channels. On the other hand, offline solutions are more computationally tractable, although they are practically infeasible due to the need of non-causal knowledge of energy arrival. In our problem setting, since the optimal online and offline solutions are equal, we find the ESP of the practical online scheme by simply formulating its offline counterpart with reduced computational complexity.

\begin{lemma}\label{lemma optimality}
The offline store-and-transmit and online pause-and-transmit protocols are optimal EH fixed rate transmission protocols. 
\end{lemma}

\begin{IEEEproof}
See Appendix \ref{app lemma optimal}.
\end{IEEEproof}

\begin{remark}
Though the intuitions behind our practical schemes and the schemes in \cite{ozel_TIT} are similar, there are some technical differences among them. The scheme of \cite{ozel_TIT} is an information-theoretic random-codebook generation that complies with instantaneous energy constraints at each channel use. However, our schemes practically order the transmitter by appropriate commands to schedule the transmission based on the harvested energy. These orders consist of the time of the transmission and the corresponding transmitted powers. In our finite-horizon transmission schemes, there are errors due to the energy shortage and our goal is to analyse the ESP performance. 

\end{remark}

\subsection{Formulation of Optimal Offline Scheme}

\begin{figure}[b!]
  \centering
  \vspace{-.1cm}
  \includegraphics[width=4in]{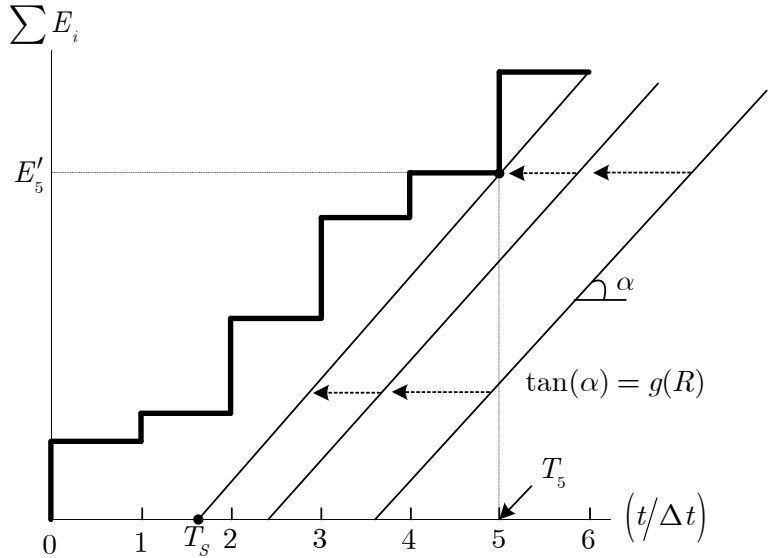}\\
  \vspace{-.1cm}
  \caption{An illustration of the energy shortage probability formulation in the offline transmission scheme.
  \vspace{-0.1cm}
  }\label{Fig2}
\end{figure}

As indicated before, in the offline store-and-transmit protocol, the Tx stores the harvested energies until it has enough energy to send at rate $R$ for the remaining transmission time in the finite-horizon. For the general relationship between power and rate defined as $p=g(r)$, the transmission start time can be formulated as (which is equal to the aggregate energy shortage time)
\begin{equation}
{T_S} = {T_{\hat i}} - \frac{{{{E'}_{\hat i}}}}{{g(R)}},
\end{equation}
where $g(R)$ is the minimum power required to send information at rate $R$ (or the slope of the transmission curve shown in Fig.~\ref{Fig2}), $T_{i}$ is the ($i\!+\!1$)-th harvesting instant, ${E'_i} = \sum\nolimits_{j = 0}^{i - 1} {{E_j}} $ and 
\begin{equation}
\hat i = \mathop {\argmax }\limits_i \left\{ {T_{i} - \frac{{{E'_i}}}{{g(R)}}} \right\}.
\end{equation}
It is worth noting that $g(R)$ can be any arbitrary monotonic function of rate. Schematically, as shown in Fig.~\ref{Fig2}, in order to find ${T_S}$, we draw a line with the slope of $g(R)$ and shift it from right to left until it is tangent to the \emph{Cumulative Harvested Energy Curve} (CHEC). The CHEC is the profile of energy transferred from ESD1 to the ESD2. A transmission curve is a line with slope $g(R)$, and its value at each time is the total consumed energy. We know that any feasible transmission curve should be placed below the CHEC to satisfy the energy causality constraint. It is easy to show that the \emph{point of tangency} of any feasible offline fixed rate transmission curve is always an EH instant (start of an epoch). The point at which a tangent line to the CHEC intersects the time axis, is called a \emph{$t$-intercept}, which may have any real value. The line associated with the maximum $t$-intercept point never intersects the CHEC elsewhere and it is the desired offline transmission curve. So, the transmission start time is the maximum of $t$-intercept points of all lines with gradient $g(R)$ tangent to CHEC. Fig.~\ref{Fig2} shows the case with $\hat i = 5$. As we see, the maximum $t$-intercept point, ${T_S}$, depends on the desired transmission rate, $R$, and the EH profile. 

Now, we are ready to formulate the ESP for a given rate of $R$. For a given EH profile of $\{E_i, i=0,...,M-1\}$, the conditional ESP is given by

\begin{IEEEeqnarray*}{rCl}
\!\!{P_{\mathrm{es}}}(R,M){|_{\left\{ {{E_i}} \right\}_{i = 0}^{M - 1}}} &\!=\!& \mathop {\max }\limits_{n,\; n=1,...,M} \frac{{{{\left\lceil {n\Delta t - \frac{{\sum\nolimits_{i = 1}^n {{E_{i - 1}}} }}{{g(R)}}} \right\rceil }^ + }}}{{M\Delta t}}
\!=\!\!{\!\left\lceil {\mathop {\max }\limits_{n = 1,...,M} \frac{n}{M}\left( {1 - \frac{{\sum\nolimits_{i = 1}^n {{E_{i - 1}}} }}{{ng(R)\Delta t}}} \right)} \right\rceil ^ +\!}.\! \yesnumber \label{offline formulation}
\end{IEEEeqnarray*}

Note that the conditional ESP may be interpreted as the Energy Shortage (ES) ratio. In fact, we find the maximum of $t$-intercept points, each associated with an EH instant as the point of tangency. If this maximum value is negative, we can start to transmit at rate $R$ right from $t=0$ and the ES ratio will be zero. Finally, averaging the ES ratio over $\{E_i\}$s, we find the ESP as ${P_{\mathrm{es}}}(R,M) = \mathbb{E}\left\{ {{P_{\mathrm{es}}}(R,M){|_{\left\{ {{E_i}} \right\}_{i = 0}^{M - 1}}}} \right\}$. 

Associated with the ESP for each finite $M$, we define the \emph{effective rate} as
\begin{equation}
\!\!{\cal R}_\mathtt{eff}\! \!=\! R (1 - {P_{\mathrm{es}}}(R,M)). 
\end{equation}

We remark that the closed-form solution of ${P_{\mathrm{es}}}(R,M)$ for any $M$ and any EH profile in not known and it seems to be complicated.
In the following, we first find the solution for some finite $M$s and then we find the asymptotic behaviour of energy shortage probability by setting $M$ to go to infinity.
\vspace{-0.1cm}
\section{Finite-horizon transmission} \label{section-finite trans}
%
In this section, we find the exact expression for the ESP and the effective rate in two cases of one epoch and two epochs, i.e., $M=1$ and $M=2$. We have assumed EH profiles with i.i.d exponentially distributed EH amounts for this analysis. Also, an upper bound on the ESP for any finite $M$ is derived. 
\subsection{Exact formula for $M=1$}
For a single EH epoch with a harvesting energy of $E_0$, the solution to the following problem
\begin{equation}
{P_{\mathrm{es}}}(R,M=1) ={\mathbb{E}_{E_0}}{\max \left\{{1-\frac{{{E_0}}}{{g(R)\Delta t}},0}\!\right\}},
\end{equation}
can be obtained as 
\begin{equation}
{P_{\mathrm{es}}}(R,M=1) = \left(1-\frac{\bar E}{\Gamma(R)}\right)+ \frac{\bar E}{\Gamma(R)}{e^ { - \frac{{\Gamma (R)}}{{\bar E}}}}, \label{Exact for M=1}
\end{equation}
where $\bar E$ is the mean energy harvesting rate and $\Gamma (R) = {g(R)\Delta t}$ is the amount of energy required to send information at rate $R$ for $\Delta t$ second.
In this way we obtain the effective rate for $M=1$ as
\begin{equation}
{\cal R}_{\mathtt{eff}}(R,M = 1) = \frac{{R\bar E}}{{\Gamma (R)}}\left( {1 - {e^{ - \frac{{\Gamma (R)}}{{\bar E}}}}} \right).
\end{equation}
\subsection{Exact formula for $M=2$}
For two energy harvesting instants (with EH values of $E_0$ and $E_1$), the problem for the ESP is
\begin{equation}
{P_{\mathrm{es}}}(R)\!=\!{\mathbb{E}_{{E_0},{E_1}}}\left[\!{\max \left\{\!{\frac{1}{2}\!-\!\frac{{{E_0}}}{{2g(R)\Delta t}},1\!-\!\frac{{{E_0} + {E_1}}}{{2g(R)\Delta t}},0}\!\right\}}\!\right].
\end{equation}
We perform a two-dimensional search in $E_0-E_1$ plane to find the solution. After some mathematical manipulations, the result will be
\begin{IEEEeqnarray}{lCl}
{P_{\mathrm{es}}}(R,M=2)& = &\left(1-\frac{\bar E}{\Gamma(R)}\right)+ \frac{\bar E}{2\Gamma(R)}e^ { - \frac{{\Gamma (R)}}{{\bar E}}}
 +\left(\frac{1}{2}+ \frac{{\bar E}}{{2\Gamma (R)}}\right)e^ { - \frac{{2\Gamma (R)}}{{\bar E}}},\label{Exact for M=2}
\end{IEEEeqnarray}
and hence,
\begin{IEEEeqnarray}{l}
{\cal R}_{\mathtt{eff}}(R,M = 2) = R\left\{ {\frac{{\bar E}}{{\Gamma (R)}} - } \right.\frac{{\bar E}}{{2\Gamma (R)}}{e^{ - \frac{{\Gamma (R)}}{{\bar E}}}} \left. { - \frac{1}{2}\left( {1 + \frac{{\bar E}}{{\Gamma (R)}}} \right){e^{ - \frac{{2\Gamma (R)}}{{\bar E}}}}} \right\},
\end{IEEEeqnarray}

where $\bar E$ is the mean of exponential distribution. Similarly, we can find the result for higher values of $M$; however, the solution is complicated. In the next section, we present the asymptotic behaviour of the ESP for infinite-horizon transmission.

\subsection{Upper Bound on the ESP for any $M$}
Now, we derive an upper bound on the ESP. 

\begin{lemma} \label{lemma M=1}
The probability of energy shortage for the single EH instant, $P_{\mathrm{es}}(R,M\!=\!1)$, is an upper bound on the ESP of the EH fixed rate transmission with any harvesting duration ($P_{\mathrm{es}}(R,M)$).
\end{lemma}
\begin{IEEEproof}
See Appendix \ref{app lemma M=1}.
\end{IEEEproof}

\section{Asymptotic Performance Behaviour} \label{section-asymptotic}
In this section, we consider the case where harvesting time goes to infinity (infinite-horizon transmission), and find closed-form solutions for asymptotic ESP and the asymptotic effective rate. 
The asymptotic ESP can be expressed as 
\begin{equation}
\!\!{\tilde P_{\mathrm{es}}}(R)\!=\!\mathop {\lim }\limits_{M \to \infty }\! \left\{\!{{\mathbb{E}_{\{ {E_i}\} }}{{\left\lceil {\mathop {\max }\limits_{n = 1,...,M} \!\frac{n}{M}\!\left( {1 - \frac{{\sum\nolimits_{i = 1}^n {{E_{i - 1}}} }}{{n\Gamma(R)}}} \right)} \right\rceil }^ + }}\! \right\}\!.
\end{equation}

In the next theorem, we find the closed-form solution of this equation.

\begin{theorem}\label{Theorem1}
The asymptotic energy shortage probability of the communication system (defined in section \ref{section-sys model}) for a fixed rate of $R$ bps is given by 

\begin{equation}
{\tilde P_{\mathrm{es}}}(R) = \left\{
\begin{IEEEeqnarraybox}[\IEEEeqnarraystrutmode
\IEEEeqnarraystrutsizeadd{2pt}{2pt}][c]{lll}
1 - \frac{{\bar E}}{\Gamma (R)}=1 - \frac{{\bar P}}{{g(R)}},{\;\; \textrm{if}\;\,R > {R_0}}\\
{0,\qquad\qquad\,\,\,\,\,\,\,\,\,\,\,\,\,\,\,\,\,\,\,\,\,\,\,\,\,\,\,\,\,\,\,\,\,\,\textrm{o.w.}}
\end{IEEEeqnarraybox}
\right.\label{theorem-equation}
\end{equation}

where $\bar P = {{\bar E} \mathord{\left/ {\vphantom {{\bar E} {\Delta t}}} \right.
 \kern-\nulldelimiterspace} {\Delta t}}$ and $R_0$ is the solution of ${g(R_0)}=\frac{{\bar E}}{\Delta t}$.
\end{theorem}

\begin{IEEEproof}
First, we define
\begin{equation}
f_n (M) = {\frac{n}{M}} \left( {1- \frac{{\sum\nolimits_{i = 1}^n {{E_{i - 1}}} }}{{{n}\Gamma (R)}}} \right),\, \forall \,n=1,...,M. \label{star-equation}
\end{equation}

In order to prove (\ref{theorem-equation}), it is enough to show that 
\begin{equation}
\!\!\mathop {\lim }\limits_{M \to \infty }\!{P_{\mathrm{es}}}(R,M)\! =\! \mathop {\lim }\limits_{M \to \infty } {\mathbb{E}_{\{ {E_i}\} }} {\left\lceil \mathop {\max }\limits_{n}{{f_n}(M)} \right\rceil ^ + } \!=\! {\left\lceil {U} \right\rceil ^ + },
\end{equation}
where $U = 1 - {{\bar E} \mathord{\left/ {\vphantom {{\bar E} {\Gamma (R)}}} \right.
 \kern-\nulldelimiterspace} {\Gamma (R)}}$.
Using (\ref{star-equation}), the mean and variance are
\begin{equation}
{\bar f_n}(M)=\mathbb{E}\left\{ {{f_n}(M)} \right\} = \frac{n}{M}\left( {1 - \frac{{\bar E}}{{\Gamma (R)}}} \right) = \frac{n}{M}U,
\end{equation}
and
\begin{equation}
\sigma _n^2(M) = \mathbb{E}\left\{ {{{\left| {{f_n}(M) - {{\bar f}_n}(M)} \right|}^2}} \right\} = \frac{n}{{{M^2}}}\frac{{\sigma_ E^2}}{{{\Gamma ^2}(R)}} = \frac{n}{{{M^2}}}\sigma _0^2,
\end{equation}
respectively, where $\sigma _E^2 = \mathbb{E}\left\{ {{{\left| {{E_i} - {{\bar E}}} \right|}^2}} \right\}$. 

Now, assume that $U>0$. This leads to $\bar f_n(M)>0, \forall n,M$. 
For $n=1,...,\sqrt{M}$, for any sufficiently small $\xi>0$ we have
\begin{IEEEeqnarray*}{rCl}
\!\!\Pr \left\{{{\ \!\! f_n(M)\!-\!U}\! \ge\! \xi\! } \right\} &\!\mathop\le\limits^{(a)}\!& \Pr \left\{ { {\ \!f_n(M)\!-\!\bar f_n(M)}\! \ge\! \xi\! } \right\}\\
&\! \le\!&\Pr \left\{\! {\left|\!{\ f_n(M)\!-\!\bar f_n(M)}\! \right| \!\ge\! \xi } \right\}\\& \!\mathop\le\limits^{(b)}\!&  \frac{{\sigma _n^2(M)}}{{{\xi ^2}}}=\frac{n}{{{M^2}}}\frac{\sigma _0^2}{{\xi ^2}} \\
&\!\mathop\le\limits^{(c)}\!&\frac{1}{{{M\sqrt{M}}}}\frac{\sigma _0^2}{{\xi ^2}},\yesnumber
\end{IEEEeqnarray*}
where (a) follows from $\bar f_n(M)<U$, (b) follows from Chebishev's inequality, and (c) follows from $n \le \sqrt{M}$. If we define ${X_i} = 1 - {{{E_{i - 1}}} \mathord{\left/ {\vphantom {{{E_{i - 1}}} {\Gamma (R)}}} \right. \kern-\nulldelimiterspace} {\Gamma (R)}}$, then $f_n(M)=\frac{{\sum\nolimits_{i = 1}^n {{X_i}} }}{M}$. For large enough values of $n$, according to the Central Limit Theorem (CLT), $f_n(M)$ is normally distributed. So for large $M$ and for $n=\sqrt{M}+1,...,M$, 
\begin{IEEEeqnarray*}{rCl}
\!\!\Pr \left\{{{\ \!\! f_n(M)\!-\!U}\! \ge\! \xi\! } \right\} &\!\le\!& \Pr \left\{ { {\ \!f_n(M)\!-\!\bar f_n(M)}\! \ge\! \xi\! } \right\}\\
&\! \mathop  = \limits^{(a)}\!&Q(\frac{\xi}{\sigma_n})\\
& \!\mathop  \le \limits^{(b)}\!& \frac{1}{2}e^{-\frac{\xi^2}{2\sigma_n^2}}=\frac{1}{2}e^{-\frac{M^2\xi^2}{2n \sigma_0^2}}  \\
&\!\mathop  \le \limits^{(c)}\!&{\left( {{e^{ - \frac{{{\xi ^2}}}{{2\sigma _0^2}}}}} \right)^n}={a^n}(\xi ), \yesnumber
\end{IEEEeqnarray*}
where (a) follows from the definition of $Q$-function, (b) follows from the Chernoff bound for $Q$-function ($Q(x)\le \frac{1}{2}e^{-\frac{x^2}{2}}, x>0$), and (c) follows from $n\le M, \forall n$.

Also, for bounded distributions, we use the Hoeffding inequality \cite{Hoeffding} to prove the theorem.  
\begin{theorem}[{\cite[Theorem 2]{Hoeffding}}]
If $X_1,X_2,...,X_n$ are independent and $\alpha_i \le X_i \le \beta_i, (i=1,2,...,n)$, then for $\lambda>0$
\begin{equation}
\Pr \left\{ {\frac{{{X_1} + {X_2} + ... + {X_n}}}{n} - \bar X \ge \lambda } \right\} \le {e^{{{ - 2{n^2}{\lambda ^2}} \mathord{\left/
 {\vphantom {{ - 2{n^2}{\lambda ^2}} {\sum\nolimits_{i = 1}^n {{{({\beta _i} - {\alpha _i})}^2}} }}} \right.
 \kern-\nulldelimiterspace} {\sum\nolimits_{i = 1}^n {{{({\beta _i} - {\alpha _i})}^2}} }}}}
\end{equation}
\end{theorem}
If we assume that $\forall i, {\beta _i} - {\alpha _i}<A$, then
\begin{equation}
\Pr \left\{ {\sum\limits_{i = 1}^n {\left( {{X_i} - \bar X} \right)}  \ge \lambda } \right\} \le {\left( {{e^{{{ - 2{\lambda ^2}} \mathord{\left/ {\vphantom {{ - 2{\lambda ^2}} A}} \right. \kern-\nulldelimiterspace} A}}}} \right)^n}.
\end{equation}
It is obvious that $a={{e^{{{ - 2{\lambda ^2}} \mathord{\left/ {\vphantom {{ - 2{\lambda ^2}} A}} \right. \kern-\nulldelimiterspace} A}}}}<1$.


The union bound can be written as

\begin{IEEEeqnarray}{rCl}
\!\!\!\Pr \left\{ {\bigcup\limits_{n = 1}^M {\left( {{f_n}(M) - {{\bar f}_M}(M) \ge \zeta } \right)} } \right\} &\le\!&\! \sum\limits_{n = 1}^M {\left( {{f_n}(M) - {{\bar f}_M}(M) \ge \zeta } \right)}\nonumber\\ 
&=& \sum\limits_{n = 1}^{\sqrt M } {\left( {{f_n}(M) - {{\bar f}_M}(M) \ge \zeta } \right)} 
\!+\!\! \sum\limits_{n = \sqrt M  + 1}^M {\!\!\left( {{f_n}(M) - {{\bar f}_M}(M) \ge \zeta } \right)} \nonumber\\
\!&\!\le\!&\! \sqrt M  \cdot \frac{1}{{M\sqrt M }}\frac{{\sigma _0^2}}{{{\zeta ^2}}} \!+\!\sum\limits_{\sqrt M  + 1}^M {{a^n}}
\le \frac{1}{{M}}\frac{{\sigma _0^2}}{{{\zeta ^2}}} \!+\!\sum\limits_{\sqrt M  + 1}^\infty  {{a^n}}\nonumber\\\!&=&\!
 \frac{1}{{M}}\frac{{\sigma _0^2}}{{{\zeta ^2}}} \!+\! \frac{{{a^{\sqrt M  + 1}}}}{{1 - a}}\! = \!\Theta (M). \yesnumber\label{longequation} 
\end{IEEEeqnarray}
Now, if we let $M \to \infty$, then $\Theta (M) \to 0$. Therefore, $\Pr \left\{{{\ \!\! \mathop {\max }\limits_{n}f_n(M)\!-\!U}\! \ge\! \xi\! } \right\} \to 0$ and 
\begin{equation}
\!\!\mathop {\lim }\limits_{M \to \infty }\!{P_{\mathrm{es}}}(R,M)\! =\! \mathop {\lim }\limits_{M \to \infty } {\mathbb{E}}[U] \!=\! U,\;\;\; U>0 \label{pos}
\end{equation}.

On the other hand, if $U\le 0$, then $\bar f_n(M)\le 0, \forall n,M$. In this case
\begin{equation}
\!\!\Pr \left\{{{\ \!\! f_n(M)\!-\!0}\! \ge\! \xi\! } \right\} \!\le\! \Pr \left\{ { {\ \!f_n(M)\!-\!\bar f_n(M)}\! \ge\! \xi\! } \right\},
\end{equation}
and similar to the case of $U> 0$, union bound states that $\Pr \left\{{{\ \!\! \mathop {\max }\limits_{n}f_n(M)\!-\!0}\! \ge\! \xi\! } \right\} \to 0$ as $M \to \infty$. Thus 
\begin{equation}
\!\!\mathop {\lim }\limits_{M \to \infty }\!{P_{\mathrm{es}}}(R,M)\! =\! \mathop {\lim }\limits_{M \to \infty } {\mathbb{E}}[U] \!=\! 0,\;\;\; U \le 0 \label{neg}.
\end{equation}
Using (\ref{pos}) and (\ref{neg}), we conclude that

\begin{equation}
\begin{array}{l}
{{\tilde P}_{\mathrm{es}}}(R) = \mathop {\lim }\limits_{M \to \infty } {P_{\mathrm{es}}}(R,M)= \mathop {\lim }\limits_{M \to \infty } {\mathbb{E}}\{ {\left\lceil U \right\rceil ^ + }\}
 = {\left\lceil U \right\rceil ^ + } = \left\{
\begin{IEEEeqnarraybox}[\IEEEeqnarraystrutmode
\IEEEeqnarraystrutsizeadd{2pt}{2pt}][c]{lll}
1 - \frac{{\bar P}}{{g(R)}},{\;\;\qquad\qquad \textrm{if}\;\,R > {R_0}}\\
{0.\qquad\qquad\,\,\,\,\,\,\,\,\,\,\,\,\,\,\,\,\,\,\,\,\,\,\,\,\textrm{o.w.}}
\end{IEEEeqnarraybox}
\right.
\end{array}
\end{equation}

\end{IEEEproof}

Now, we find ${R_0}$ by solving the following equation
\begin{equation}
U (R_0) = 1 - \frac{{\bar P}}{{g(R_0)}} = 0.
\end{equation}
For Shannon power-rate formula, $g(R)=g_{sh}(R) ={{2^{2R}} - 1}$ (assuming ideal system with hardware power consumption of zero), this leads to ${R_0} = \frac{1}{2}{\log _2}(1 + \bar P).$ This means that if we send information at any fixed rate lower than ${R_0}$, the ESP asymptotically will be zero. Interestingly, this is the capacity formula for AWGN channel with stochastic energy harvesting derived in \cite{ozel_TIT}.

\begin{remark} \label{remark general g}
Our results hold for any arbitrary monotonic power consumption formula, $g(R)$. Thus, they are also applicable for the scenarios, where a typical node consumes significant energy for sensing, data processing, on/off switching in addition to the energy used for transmission. We consider such an example in section \ref{section-numeric result}, in which a node's energy consumption is limited to non-transmission consumption and the power consumption for data-transmission is negligible. 
\end{remark}

\begin{corollary}
We can asymptotically achieve the capacity of a AWGN channel under average power constraint $\bar{P}$, i.e., $\mathcal{C}_{\bar{P}}=\frac{1}{2}\log(1+\bar P)$, in the AWGN channel with a fixed rate EH-Tx (where average EH rate is $\bar{P}$).
This means that the aggregate energy shortage time will be vanished by making $M$ large enough, if the fixed transmission rate is not larger than $\mathcal{C}_{\bar{P}}$.  
      
\end{corollary}


\begin{lemma}
The asymptotic ESP for infinite-horizon transmission is a lower bound for the ESP of any finite-horizon transmission.
\end{lemma}

\begin{IEEEproof}
According to the earlier definition, $f(M)={P_{\mathrm{es}}}(R,M){|_{\left\{ {{E_i}} \right\}_{i = 0}^{M - 1}}}$, it is easy to show that
\begin{equation}
f(M)\!=\!\max\left\{\!\frac{M-1}{M}f(M-1)\!,\!{\left\lceil\!{1-\frac{{{{\bar E}_{{M}}}}}{{\Gamma (R)}}}\!\right\rceil^+}\!\right\}\!\ge\!{\left\lceil\!{1-\frac{{{{\bar E}_{{M}}}}}{{\Gamma (R)}}}\!\right\rceil^+}
\end{equation}
where ${\bar E_{{M}}} = {{\sum\nolimits_{i = 1}^{{M}} {{E_{i - 1}}} } \mathord{\left/
 {\vphantom {{\sum\nolimits_{i = 1}^{{M}} {{E_{i - 1}}} } {{M}}}} \right.
 \kern-\nulldelimiterspace} {{M}}}$. By applying the expectation operation to both sides of the above inequality, and using the \emph{Jensen's inequality} (as $h(\mathbf{x})={\left\lceil \mathbf{x} \right\rceil ^ + }$ is a convex function), we have
 \begin{equation}
\mathbb{E}\left \{{\left\lceil\!{1-\frac{{{{\bar E}_{{M}}}}}{{\Gamma (R)}}}\!\right\rceil^+} \right \}\ge {\left\lceil\!{1-\frac{\mathbb{E}\{{{{\bar E}_{{M}}}}\}}{{\Gamma (R)}}}\!\right\rceil^+}={\left\lceil\!{1-\frac{{\bar E}}{{\Gamma (R)}}}\!\right\rceil^+}.
 \end{equation}
 
 We conclude that
 \begin{equation}
 {P_{\mathrm{es}}}(R,M)\ge{\tilde P_{\mathrm{es}}}(R).
 \end{equation}
Since this equation is held for any value of $M$, the lemma is proved.  
\end{IEEEproof}

Now, we derive the asymptotic effective rate as follows:
\begin{equation}
\begin{array}{l}
\tilde {\cal R}_{\mathtt{eff}} = R\left( {1 - {{\tilde P}_{\mathrm{es}}}(R)} \right) = \left\{ {\begin{array}{*{20}{c}}
{\frac{{\bar PR}}{{({2^{2R}} - 1)}},\,\,\,\,\,\,\,\,\,\,\,\,\,\,\,{\rm{if}}\,\,\,R > {R_0}}\\
{R,\,\,\,\,\,\,\,\,\,\,\,\,\,\,\,\,\,\,\,\,\,\,\,\,\,\,\,\,\,\,\,\,{\rm{otherwise}}}
\end{array}} \right.
 = \frac{{({2^{2{R_0}}} - 1)R}}{{({2^{2R}} - 1)}}{U_0}(R - {R_0}) + R{U_0}({R_0} - R)
\end{array}
\end{equation}
where ${U_0}(.)$ is the unit step function.
As expected, the maximum achievable $\tilde {\cal R}_{\mathtt{eff}}$ is the capacity $\mathcal{C}_{\bar P}=R_0$.

\begin{figure}[t]
  \centering
  \vspace{-.1cm}
  \includegraphics[width=4.5in]{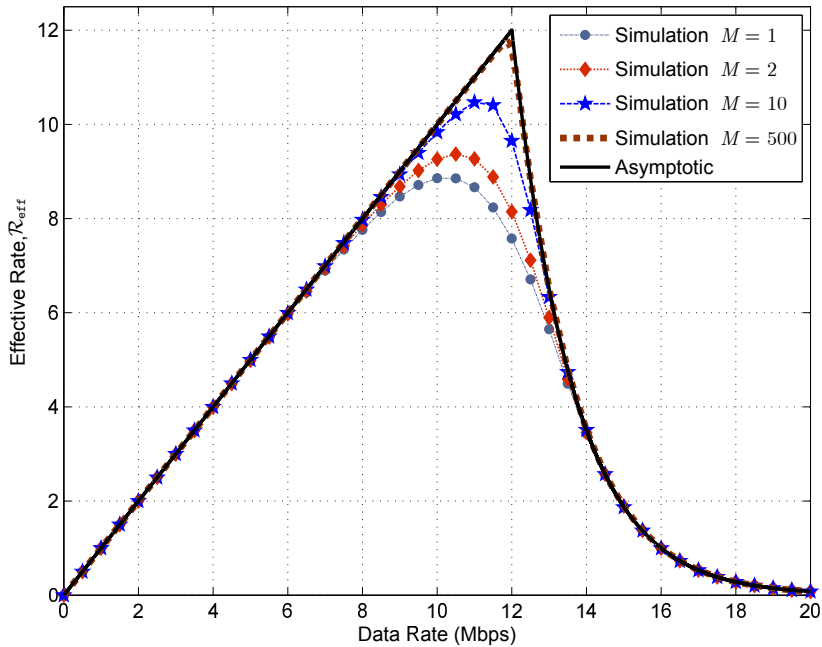}\\
  \vspace{-.1cm}
  \caption{Comparison between the simulation results for the effective rate, ${\cal R_{\mathtt{eff}}}$, for different values of $M$, and its asymptotic value for infinite-horizon transmission, $\tilde {\cal R}_{\mathtt{eff}}$. Results are presented for ${\cal C} = 12$Mbps.
  \vspace{-0.1cm}
  }\label{throughput}
\end{figure}

Fig.~\ref{throughput} shows the $\tilde {\cal R}_{\mathtt{eff}}$ in terms of $R$ for ${R_0} = 12\,\mathrm{Mbps}$. Also, the effective rate for different finite transmission durations are shown in this figure. We are interested in the value of $R$ that maximizes the ${\cal R}_{\mathtt{eff}}$ for any $M$. Therefore, for any finite transmission duration, we can provide the system designer with the maximum achievable effective rate that should be used for data transmission. For example, Fig. ~\ref{throughput} shows that the maximum achievable effective rate is ${\cal R_{\mathtt{eff}}}=8.869$ for $M=1$ (${\cal R_{\mathtt{eff}}}=0.739\,\mathcal{C}$), which is occurred at $R=10.21\,\mathrm{Mbps}$. Thus, we should send data at the fixed rate of $R=10.21\,\mathrm{Mbps}$ for $M=1$, whenever we have sufficient energy. 

As can be seen in this figure, the maximum ${\cal R_{\mathtt{eff}}}$ and its associated $R$ decreases as $M$ decreases. Also, the capacity is only asymptotically achievable.

\section{Outage Analysis in Fading Channels} \label{section-fading}
In this section, we analyse the outage performance of EH fixed rate communications in the \emph{fading} channels with no CSIT. For $M$ energy harvesting epochs, we assume that there are $N$ i.i.d. fade levels, as depicted in Fig.~\ref{fading time representation}. Now, channel fading as well as energy shortage can cause outage. We find the corresponding theoretical results for the outage probability of fading channel without any knowledge about CSI at the Tx.
For fading channels, the instantaneous power ($p_i$) in terms of the fixed rate of $R$ can be expressed as ${p_i} = g(R,{\gamma _i}) = {{{g(R)} \mathord{\left/ {\vphantom {{g({r_i})} \gamma }} \right. \kern-\nulldelimiterspace} \gamma }_i}$, where ${\gamma}_i$ is the instantaneous channel power gain defined as ${\left| {{h_i}} \right|^2}$.
In the case of no CSIT, the Tx in the online scheme may transmit whenever the sufficient energy for transmission at a predetermined rate of $R$ is available. Moreover, transmitted information may be lost due to the fading (second source of outage). 

Even though there is no CSIT, we may use the distribution knowledge of the channel (CDIT) to improve the performance. 
In fact, we use the CDIT to find an optimal threshold for channel gain. This threshold is used for decision about transmission initiation. We derive these optimal values for any transmission rate and any transmission period, which can be calculated and tabulated before communication commences.   

First note that the effective rate in the fading environment with CDIT can be written as
\begin{equation}
\mathcal{R}_{\mathtt{eff,F}}=\mathcal{R}_{\mathtt{eff}}(1-P_{\mathrm{ch}}({\gamma _{\mathtt{thr}}}))=R(1-P_{\mathrm{es}}(R,M))(1-P_{\mathrm{ch}}({\gamma _{\mathtt{thr}}})),
\end{equation}
where ${\gamma _{\mathtt{thr}}}$ is the threshold channel power gain, above which the transmission occurs and below which the transmission is halted. Also, ${P_{\mathrm{ch}}}$ is the outage probability due to the channel fading and (for Rayleigh fading channels) is given by 

\begin{equation}
\begin{array}{l}
{P_{\mathrm{ch}}}({\gamma _{\mathtt{thr}}}) = \Pr \left\{ {\gamma  < {\gamma _{\mathtt{thr}}}} \right\} = {F_\gamma }({\gamma _{\mathtt{thr}}}) = \int_0^{{\gamma _{\mathtt{thr}}}} {\frac{1}{{{\gamma _0}}}} {e^{ - \frac{\gamma }{{{\gamma _0}}}}} = 1 - {e^{ - \frac{{{\gamma _{\mathtt{thr}}}}}{{{\gamma _0}}}}}.
\end{array}
\end{equation}

We define the probability of outage of EH communications in the fading channels with CDIT as 
\begin{equation}
\mathcal{R}_{\mathtt{eff,F}} \triangleq R(1-{P_{\mathrm{out}}}(R,M,{\gamma _{\mathtt{thr}}})).
\end{equation}
This leads to the following expression for the outage probability
\begin{equation}
{P_{\mathrm{out}}}(R,M,{\gamma _{\mathtt{thr}}}) = 1 - \left( {1 - {P_{\mathrm{es}}}(R,M)} \right) \times \left( {1 - {P_{\mathrm{ch}}}({\gamma _{\mathtt{thr}}})} \right).
\end{equation}

The optimal thresholds are the solution of the following optimization problem

\begin{equation}
{\gamma _{\mathtt{thr}}^*}(R,M) = \mathop {\argmin }\limits_{{\gamma _{\mathtt{thr}}}} {P_{\mathrm{out}}}(R,M,{\gamma _{\mathtt{thr}}}) \label{optimization problem}.
\end{equation}

\begin{figure}[t!]
  \centering
  \vspace{-.1cm}
  \includegraphics[width=4.5in]{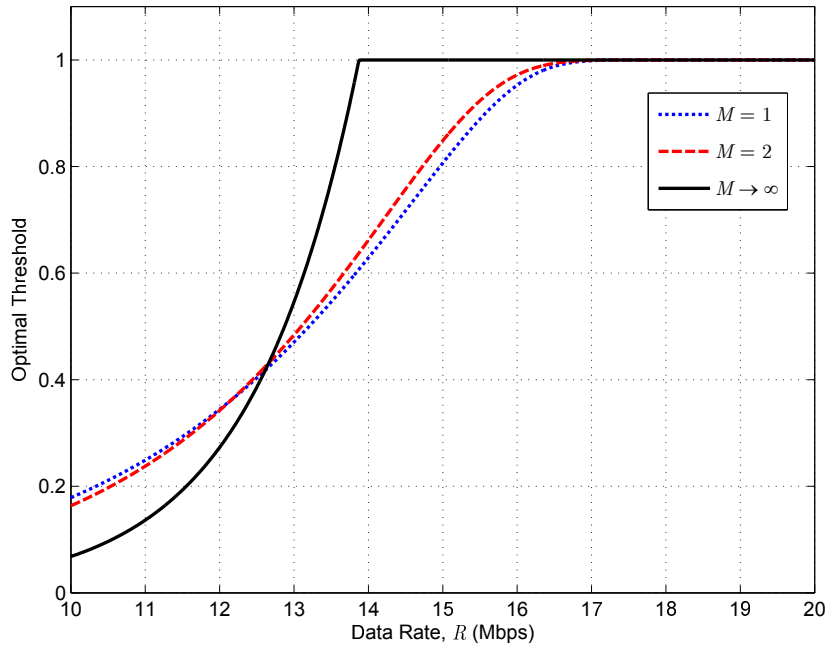}\\
  \vspace{-.1cm}
  \caption{Optimal channel power gain threshold for transmission initiation in the online scheme versus data rate. Optimization is performed for normalized Rayleigh fading channels with $M=1$ and $M=2$ and infinite-horizon transmission $M \to \infty$.
  \vspace{-0.1cm}
  }\label{Optimal_threshold}
\end{figure}

\begin{figure}[t]
  \centering
  \vspace{-.1cm}
  \includegraphics[width=4.5in]{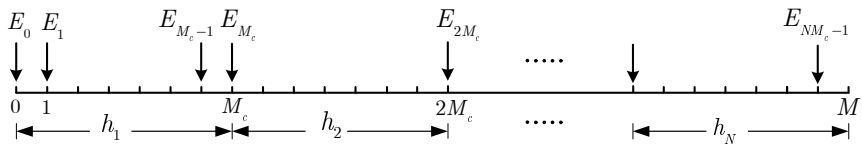}\\
  \vspace{-.1cm}
  \caption{Temporal representation of EH in fading channels.
  \vspace{-0.1cm}
  }\label{fading time representation}
\end{figure}

It is worth noting that these optimal threshold values can be calculated in an offline mode by using only the CDIT information. Then, in the beginning of each epoch, knowing its harvested energy, this threshold is utilized for calculating the power required for transmission in the determined rate (${p} = g({R},{\gamma _{\mathtt{thr}}^*}) = {{{g({R})} \mathord{\left/ {\vphantom {{g({R})} \gamma }} \right. \kern-\nulldelimiterspace} \gamma }_{\mathtt{thr}}^*}(R,M)$).
Fig. \ref{Optimal_threshold} portrays the corresponding optimal threshold values for $M=1$, $M=2$ and $M \to \infty$ in terms of data rate.  

%
%

Theoretical expressions for the outage probability of finite-horizon transmissions ($M=1$ and $M=2$) in the fading channels with CDIT, using (\ref{Exact for M=1}) and (\ref{Exact for M=2}), are given by

\begin{equation}
{P_{\mathrm{out}}}(R,M = 1,{\gamma _{\mathtt{thr}}}) = 1 - K{\gamma _{\mathtt{thr}}}{e^{ - {\gamma _{\mathtt{thr}}}}}\left( {1 - {e^{ - \frac{1}{{K{\gamma _{\mathtt{thr}}}}}}}} \right),
\end{equation}

\begin{equation}
\begin{array}{l}
{P_{\mathrm{out}}}(R,M = 2,{\gamma _{\mathtt{thr}}}) = 1 - {e^{ - {\gamma _{\mathtt{thr}}}}}\left( {K{\gamma _{\mathtt{thr}}} - \frac{{K{\gamma _{\mathtt{thr}}}}}{2}{e^{ - \frac{1}{{K{\gamma _{\mathtt{thr}}}}}}}} \right.\left. { - \frac{1}{2}(1 + K{\gamma _{\mathtt{thr}}}){e^{ - \frac{2}{{K{\gamma _{\mathtt{thr}}}}}}}} \right),
\end{array}
\end{equation}
where $K=\frac{\bar E}{\Gamma(R)}$.

Finally, using the result of theorem \ref{Theorem1}, the outage probability of infinite-horizon transmission in the Rayleigh fading channel with CDIT is given by 

\begin{equation}
\begin{array}{l}
{P_{\mathrm{out}}}(R,M \to \infty ,{\gamma _{\mathtt{thr}}}) = 1 - {e^{ - {\gamma _{\mathtt{thr}}}}}\left( {1 - {{\left[ {1 - K{\gamma _{\mathtt{thr}}}} \right]}^ + }} \right) = \left\{ {\begin{array}{*{20}{c}}
{1 - K{\gamma _{\mathtt{thr}}}{e^{ - {\gamma _{\mathtt{thr}}}}}\,\,\,\,\,\,\,\,\,\,\,\,\,\,K < \frac{1}{{{\gamma _{\mathtt{thr}}}}}}\\
{1 - {e^{ - {\gamma _{\mathtt{thr}}}}}\,\,\,\,\,\,\,\,\,\,\,\,\,\,\,\,\,\,\,\,\,\,\,\,\,\,\,\,\,\,\,\,\,\,\,o.w.}
\end{array}} \right.
\end{array}
\end{equation}

In the section \ref{section-numeric result}, we show the performance gain obtained by exploiting optimal threshold values for the channel power gains.

\subsection{Online and Offline Schemes}
In the fading channels with no CSIT, similar to the AWGN channels, the optimal online and offline EH fixed rate transmission schemes are equal. Here, in online scheme, the Tx transmits data at a fixed predetermined rate whenever it has enough energy (same as the AWGN channel, except the use of optimal channel power gain threshold for transmission initiation that is a priori available) and pause the transmission, otherwise. In offline scheme, using optimal threshold values obtained in (\ref{optimization problem}), the Tx starts to transmit when the transmission at the fixed predetermined rate is probable for the remaining time period. The key point for deducing the equivalence of online and offline schemes is that knowing ${\gamma _{\mathtt{thr}}^*}$ for any $R$ and $M$, the ${P_{\mathrm{es}}}$  for online and the offline schemes are the same. 
  
We formulate the outage probability in the online scheme (using the results of offline formulation) as follows
\begin{equation}
\begin{array}{l}
{P_{\mathrm{out}}}(R,M) = {\mathbb{E}_{\{ {E_i}\} }}\frac{1}{M}\sum\limits_{i = 0}^{M - 1} {\left\{ {{{\left\lceil {1 - \frac{{{E_i}{\gamma _{\mathtt{thr}}}(R,M)}}{{\Gamma (R)}}} \right\rceil }^ + }} \right.} \left. { + \left( {1 - {{\left\lceil {1 - \frac{{{E_i}{\gamma _{\mathtt{thr}}}(R,M)}}{{\Gamma (R)}}} \right\rceil }^ + }} \right) \times {1_{\left\{ {{\gamma _i} < {\gamma _{\mathtt{thr}}}} \right\}}}} \right\}
\end{array}
\end{equation}
where the first term in the summation is the no-transmit time due to the energy shortage and the second term is due to the channel fading. We also have
\begin{equation}
{1_{\left\{ {{\gamma _i} < {\gamma _{\mathtt{thr}}}} \right\}}} = \left\{ {\begin{array}{*{20}{c}}
{1\,\,\,\,\,\,\,\,\,\,\,\,\,\,{\rm{if}}\,\,\,\,{\gamma _i} < {\gamma _{\mathtt{thr}}}}\\
{0\,\,\,\,\,\,\,\,\,\,\,\,\,\,\,\,\,\,\,\,\,\,\,\,\,\,\,\,\,\,\,{\rm{o}}{\rm{.w}}{\rm{.}}}
\end{array}} \right.
\end{equation}

\vspace{-0.3cm}
Note that we do not average over fade levels, since it is impossible if there is no CSIT. Numerical results illustrated in the next section have shown a good agreement between the performance of the online scheme and the theoretical formulas derived in this section.

\begin{figure}[t!]
  \centering
  \vspace{-.1cm}
  \includegraphics[width=4.5in]{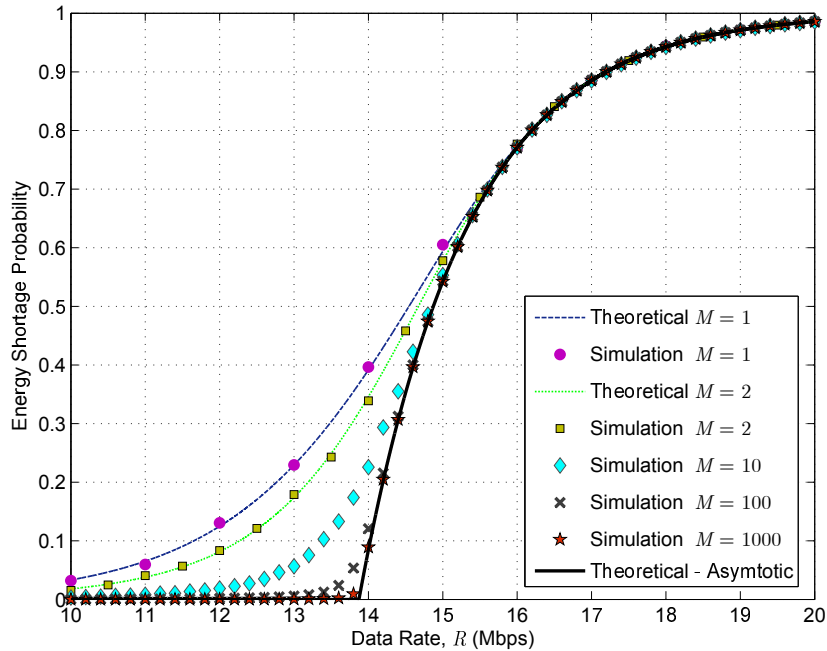}\\
  \vspace{-.1cm}
  \caption{Comparison between simulation results and theoretical values obtained from exact energy shortage probability (ESP) expressions for $M=1$ and $M=2$ and asymptotic behaviour of the ESP for infinite-horizon transmission. Results are presented for $\bar E = 15\,{\rm{mJ}}$ and different values of $M$.
  \vspace{-0.1cm}
  }\label{outage2}
\end{figure}

\section{Numerical Results and Discussions} \label{section-numeric result}
In this section, we verify our theoretical results using numerical simulations in both AWGN and fading channels. We consider a band-limited AWGN channel with noise power spectral density of ${N_0} = {10^{ - 19}}\,\,{\rm{W/Hz}}$ and bandwidth of $W = 1\,\,{\rm{MHz}}$. The distance between the Tx and the Rx is assumed to be 1 Km and path loss is $\psi=70$ dB (typical values used in some EH literatures, e.g. \cite{yang}). Substituting these parameters into the Shannon capacity formula results in $R = W{\log _2}(1 + p\psi/{N_0}W) = {\log _2}(1 + p/{10^{ - 6}})\,\,{\rm{Mbps}}$ for transmission in an AWGN channel. For the fading case, we consider normalized i.i.d. slow Rayleigh fading. For the energy harvesting profile, without loss of generality, we assume that $\Delta t = 1$ and energy harvesting values (except for Fig.~\ref{Process_Power_Poisson}) come from an exponential distribution with mean energy of $\bar E = 15\,{\rm{mJ}}$. For Fig.~\ref{Process_Power_Poisson}, we use Poisson distribution to generate random EH samples with the same mean. As mentioned in Remark \ref{remark general g}, our results can be applied to any monotonic function for the power-rate relationship. In Fig.~\ref{Process_Power_Poisson}, we use $p=g(R)=k_0+k_1R$ ($k_0$ and $k_1$ are constant values), which may account for power consumption due to sensing, data processing and on/off switching of an idle listening/monitoring node. Note that, in all other figures, Shannon formula is used. 
\begin{figure}[t!]
  \centering
  \vspace{-.1cm}
  \includegraphics[width=4.5in]{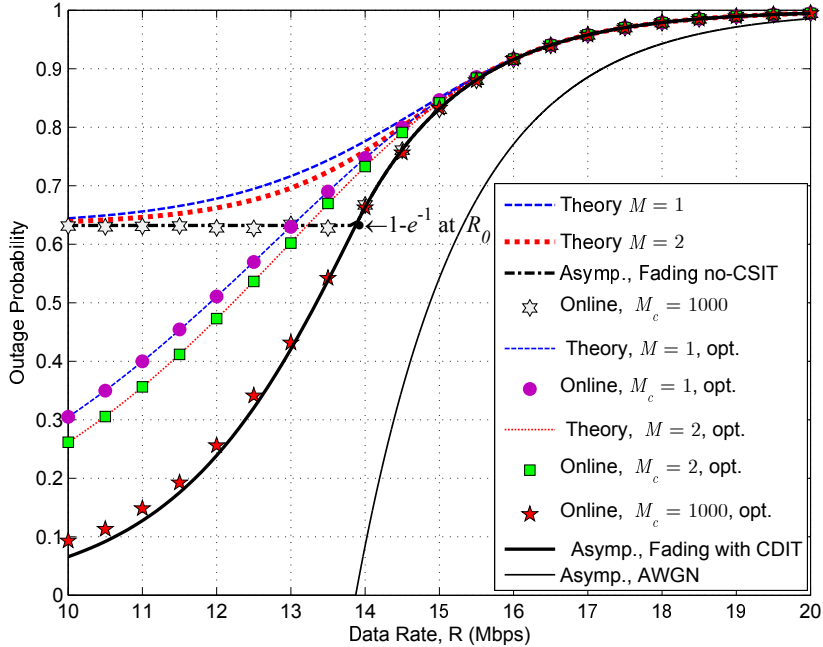}\\
  \vspace{-.1cm}
  \caption{Comparison among outage probability of fading channels with no channel side information at the transmitter; optimized outage probability, obtained using channel distribution information. The asymptotic ESP curve of AWGN channel is also included for comparison. 
  \vspace{-0.1cm}
  }\label{fading_new}
\end{figure}

Fig.~\ref{outage2} shows the effect of transmission duration (or harvesting period, i.e., $T=M\Delta t$) on the energy shortage performance of our system. Precisely, the accuracy of \emph{asymptotic} theoretical formula is investigated by increasing $M$. In this regard, the ESP is shown versus data rate for mean energy harvesting rate of $15 \textrm{mJ}$. Also, exact closed-form expressions for the ESP of $M=1$ and $M=2$ is shown. Simulation results validate theoretical formulas for these two finite-horizon transmissions. As we see, the simulation results reach the asymptotic ESP curve by increasing the number of EH instants. This confirms our theoretic results. Also, the asymptotic ESP forms a lower bound to the ESP of any finite-horizon transmission. This lower bound is tight for high data rates (approximately greater than 16Mbps) and large enough values of $M$.

The fading case with no CSIT is studied in Fig.~\ref{fading_new}, wherein the effects of channel fading on the deterioration of outage probability for finite and infinite-horizon transmissions are shown using Monte-Carlo simulations and theoretical results of the previous sections. Theoretical curves of this figure consist of ($a1$) asymptotic ESP for AWGN channel, ($a2$) theoretical curves for finite-horizon transmission ($M=1,\,2$), ($a3$) theoretical curves for infinite-horizon transmission (asymptotic), and ($a4$) the previous case (i.e., ($a3$)) where the threshold channel power gains for transmission initiation is optimized using the CDIT. Also, the simulation curves consist of ($s1$) the online fixed rate transmission scheme with different coherence time of the channel ($M_c=1,2,1000$), and ($s2$) the online scheme that uses optimal threshold for transmission initiation ($M_c=1000$). For fading channels, we are interested in the study of finite-horizon transmission that contains several fade levels, in order to study the effects of fading. We study finite-horizon transmission of length $T=10000$ and coherence times of $T_c=M_c\Delta t$. As can be seen, the theoretical results are in agreement with the simulation results. For example, the difference between the theoretical asymptotic curve with CDIT and the online scheme with $M_c=1000$ is lower than $0.03$ at $R=10$Mbps and goes to zero as data rate increases.
We observe that at the lower data rates, the threshold optimization significantly improves the outage performance of the system compared to the case of no CDIT. However, at the rates higher than $R=R_0$, where $R_0$ satisfies $g(R_0)=\bar P$, the CDI provides no gain. This is expected as according to Fig.~\ref{Optimal_threshold}, we have ${\gamma _{\mathtt{thr}}^*}=1$ for $R \ge R_0$. Note that in the case of no CDIT, we use fixed threshold for all data rate, i.e. ${\gamma _{\mathtt{thr}}^*}=1,\,\forall R$. This is suboptimal for lower data rates. 

\begin{figure}[t!]
  \centering
  \vspace{-.1cm}
  \includegraphics[width=4.5in]{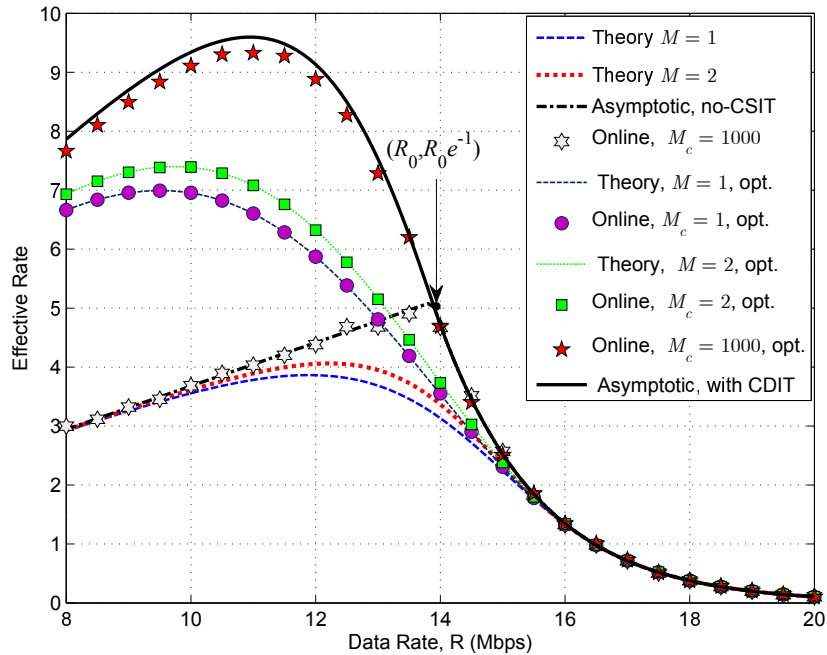}\\
  \vspace{-.1cm}
  \caption{Comparison between effective rate of fading channels, i.e., $\mathcal{R}_{\mathtt{eff,F}}$, without channel side information at the transmitter, and optimized values, obtained using channel distribution information at the transmitter. 
  \vspace{-0.1cm}
  }\label{fading_throuput}
\end{figure}

Fig.~\ref{fading_throuput} shows the corresponding curves of Fig.~\ref{fading_new} for the effective rates, with the same simulation settings. For any finite $M$, one can use these type of curves to design a transmitting data rate, which achieves the maximum effective rate. For example, the maximum achievable effective rate for $M=1$, $M=2$ and $M \to \infty$ are 7 Mbps, 7.4 Mbps and 9.6 Mbps, respectively. These values are associated with the fixed data rates of 9.5 Mbps, 9.7 Mbps, and 10.9 Mbps, respectively. 

In Fig.~\ref{Process_Power_Poisson}, we consider a quite different model for power consumption. In this model, a node consumes significant power for sensing, data processing, on/off switching and powering on baseband Digital Signal Processing (DSP) and front-end circuits. We model power consumption, without loss of generality, as $p=g(R)=k_0+k_1R$, with $k_0=1$mJ/sec and $k_1=1$nanoJ/bit, and we study the ESP of system with i.i.d. Poisson distributed EH values with mean $\bar E = 15\,{\rm{mJ}}$. Similar to Fig.~\ref{outage2}, this figure illustrates that by increasing the number of EH instants, the simulation results reach the asymptotic ESP curve. This confirms our theoretic results presented in theorem \ref{Theorem1}. Again, the asymptotic ESP forms a lower bound to the ESP of any finite-horizon transmission. 

\begin{figure}[t!]
  \centering
  \vspace{-.1cm}
  \includegraphics[width=4.5in]{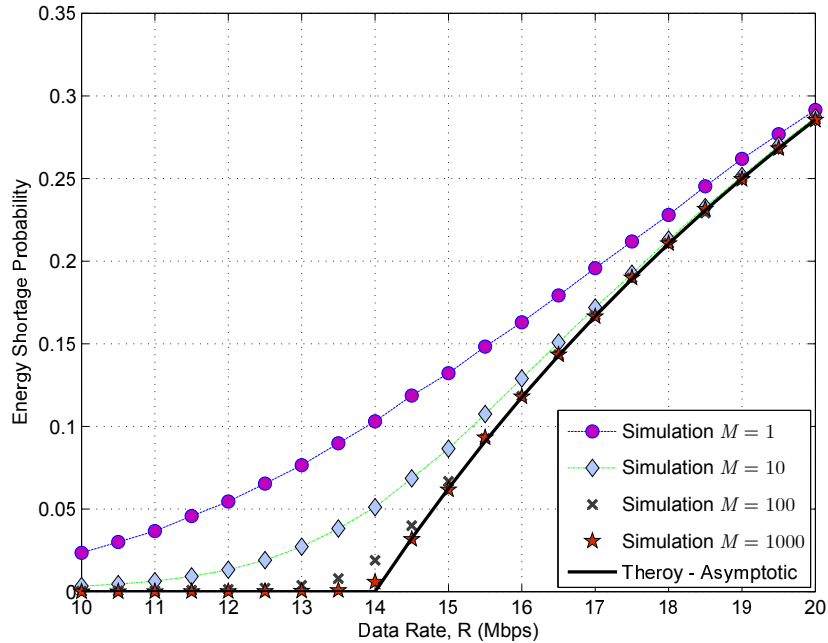}\\
  \vspace{-.1cm}
  \caption{Simulation results and asymptotic behaviour of the ESP for infinite-horizon transmission for a node whose power consumption is modelled as $p=g(R)=k_0+k_1R$, where $k_0=1$mJ/sec and $k_1=1$nanoJ/bit.
  \vspace{-0.1cm}
  }\label{Process_Power_Poisson}
\end{figure}




\section{Conclusion} \label{section-conclusion}
In this paper, we considered the energy shortage probability analysis of Energy Harvesting-capable nodes with fixed rate transmission in AWGN channels. We presented: (1) the online and offline transmission protocols that lead to the same energy shortage probability, (2) the exact closed-form solutions for finite-horizon trasmission protocols with one and two energy harvesting instants, (3) the closed-form expressions for the asymptotically tight lower bound on the energy shortage probability, and (4) the closed-form expressions for the asymptotically tight upper bound on the effective rate. Our analysis showed that the capacity of a AWGN channel under average power constraint is asymptotically achievable in the AWGN channel with a fixed rate EH transmitter.

For fading channels, we considered outage probability and the effective rate as the performance metrics, and proposed the online and offline transmission protocols with the same performance. For these metrics, we derived the closed-form exact-solutions and bounds for finite and infinite-horizon transmissions, respectively. Also, in the case of having no channel state information at the Tx, we proposed an online scheme that optimized the channel to noise ratio threshold for transmission initiation. We observed that optimizing the channel power gain threshold improves the outage performance of the system significantly compared to the case of no CDIT at lower data rates. Finding closed-form solutions for any finite transmission time and any EH profile is an interesting open problem, which is part of our future work. 


\appendices

\section{Proof of Lemma \ref{lemma optimality}} \label{app lemma optimal}
\begin{figure}[t!]
  \centering
  \vspace{-.1cm}
  \includegraphics[width=4in]{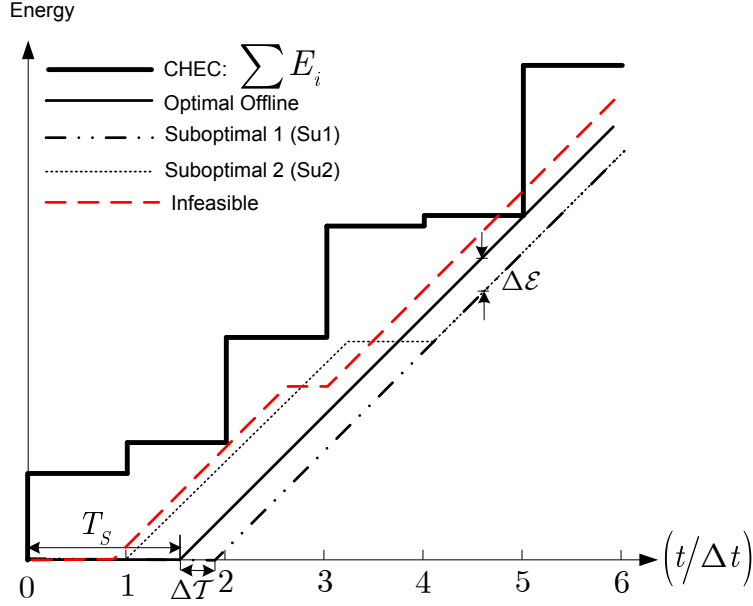}\\
  \vspace{-.1cm}
  \caption{An optimal offline store-and-transmit scheme, two suboptimal schemes and an infeasible transmission scheme are compared.
  \vspace{-0.1cm}
  }\label{Proof_Appendix}
\end{figure}

We have shown in Fig.~\ref{Fig1} (and the related discussions) that the same ESP performance can be achieved by both the online pause-and-transmit and the offline store-and-transmit protocols.
Therefore, it is enough to show the optimality of the offline store-and-transmit protocol for EH fixed rate transmission. 
Referring to Fig.~\ref{Proof_Appendix}, any transmission curve that is below the optimal curve at some time instant by the amount of $\Delta \mathcal{E}$, leads to excess shortage time of $\Delta {\cal T} = \frac{{\Delta {\cal E}}}{{g(R)}}$. Thus, it is suboptimal (like Su1 and Su2 in Fig.~\ref{Proof_Appendix}).
On the other hand, any transmission curve that is completely above the optimal transmission curve, although has lower shortage time, is infeasible. This is due to the fact that the optimal transmission curve is tangent to the cumulative harvested energy curve (CHEC) and hence any curve above this optimal curve exceeds the CHEC and violates the energy causality constraint. 
This means that any feasible transmission protocol with only two rates, $0$ and $R$, leads to higher ES ratio for any EH realization. 
This completes the proof.
 
\section{Proof of Lemma \ref{lemma M=1}} \label{app lemma M=1}

First, we prove the following lemma by defining $\mathsf{P}(M)=P_{\mathrm{es}}(R,M)=\mathbb{E}\{f(M)\}$.

\begin{lemma}\label{Lem_upper}
We have the following inequality
\begin{equation}
\mathsf{P}(M) < \mathsf{P}(M_1)\cdot\frac{M_1}{M}+\mathsf{P}(M_2)\cdot\frac{M_2}{M},
\end{equation}
for any $M_1$ and $M_2$ where $M={M_1}+{M_2}$.
\end{lemma}

\begin{IEEEproof}
We divide the total transmission period, $M$, into two disjoint intervals, $M_1$ and $M_2$, and find the ESP for each of them, separately (see Fig. \ref{lower bound}).

\begin{figure}[t]
  \centering
  \vspace{-.1cm}
  \includegraphics[width=4in]{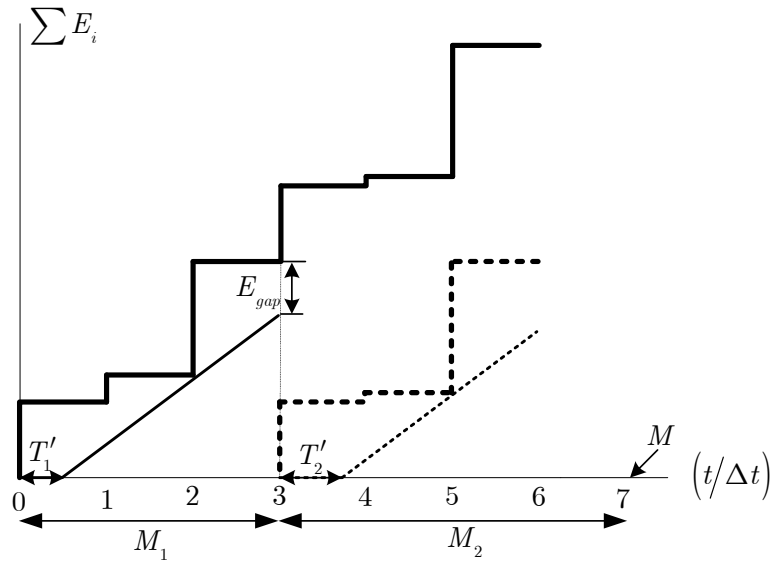}\\
  \vspace{-.1cm}
  \caption{Partitioning the total transmission time into two disjoint intervals and finding the energy shortage probability of each interval, separately.
  \vspace{-0.1cm}
  }\label{lower bound}
\end{figure}

We lose some energy ($E_{gap}$) by such division, which leads to the increment in the energy shortage time. We can easily show that the aggregate energy shortage time can be given by $T_S \le T_{S,1}+T_{S,2}$ (the equality occurs when we have $E_{gap}=0$).

For any EH pattern we have
\begin{equation}
\frac{T_S}{M} \le \frac{T_{S,1}}{M}+\frac{T_{S,2}}{M}.
\end{equation}

Since always we have $E_{gap} \ge 0$, so $\mathbb{E}\{E_{gap}\} > 0$; therefore, by averaging over EH values (different EH patterns), we can write
\begin{equation}
\mathsf{P}(M) \!<\!\mathbb{E}\{\frac{T_{S,1}}{M_1}\}\frac{M_1}{M}+\mathbb{E}\{\frac{T_{S,2}}{M_2}\}\frac{M_2}{M}\!=\! \mathsf{P}(M_1)\cdot\frac{M_1}{M}+\mathsf{P}(M_2)\cdot\frac{M_2}{M}.
\end{equation}
The proof is completed.
\end{IEEEproof}
Now, we prove the following corollary
\begin{corollary}\label{Cor_upper}
For any $K>0$, we have $\mathsf{P}(2K)<\mathsf{P}(K)$.
\end{corollary}
\begin{IEEEproof}
This can be verified by setting $M_1=M_2=K$ in Lemma \ref{Lem_upper}.
\end{IEEEproof}

From corollary \ref{Cor_upper} we have
\begin{equation}
\mathsf{P}(2^K)<\mathsf{P}(2^{K-1})<\cdots<\mathsf{P}(2)<\mathsf{P}(1), \forall K \in \mathbb{N}.
\end{equation}
Also, lemma \ref{Lem_upper} states that
\begin{equation}
\mathsf{P}(2^K+1)<\frac{2^K}{2^K+1}\mathsf{P}(2^K)+\frac{1}{2^K+1}\mathsf{P}(1)<\mathsf{P}(1), \forall K \in \mathbb{N}
\end{equation}
Similarly for $i=2,...$ and using induction on $i$
\begin{equation}
\begin{array}{l}
\!\!\!\!\!\mathsf{P}(2^K+i)<\frac{2^K+i-1}{2^K+i}\mathsf{P}(2^K+i-1)+\frac{1}{2^K+i}\mathsf{P}(1)<\mathsf{P}(1),\;\; \forall K \in \mathbb{N}
\end{array}
\end{equation}
We conclude that 
\begin{equation}
\mathsf{P}(M)<\mathsf{P}(1), \forall M \in \mathbb{N}. 
\end{equation}
This completes the proof.

\bibliographystyle{IEEEtran}
\vspace{-0.1cm}
\bibliography{reference}

\begin{thebibliography}{10}
\providecommand{\url}[1]{#1}
\csname url@samestyle\endcsname
\providecommand{\newblock}{\relax}
\providecommand{\bibinfo}[2]{#2}
\providecommand{\BIBentrySTDinterwordspacing}{\spaceskip=0pt\relax}
\providecommand{\BIBentryALTinterwordstretchfactor}{4}
\providecommand{\BIBentryALTinterwordspacing}{\spaceskip=\fontdimen2\font plus
\BIBentryALTinterwordstretchfactor\fontdimen3\font minus
  \fontdimen4\font\relax}
\providecommand{\BIBforeignlanguage}[2]{{%
\expandafter\ifx\csname l@#1\endcsname\relax
\typeout{** WARNING: IEEEtran.bst: No hyphenation pattern has been}%
\typeout{** loaded for the language `#1'. Using the pattern for}%
\typeout{** the default language instead.}%
\else
\language=\csname l@#1\endcsname
\fi
#2}}
\providecommand{\BIBdecl}{\relax}
\BIBdecl

\bibitem{sudev}
S.~Sudevalayam and P.~Kulkarni, ``Energy harvesting sensor nodes: Survey and
  implications,'' \emph{IEEE Commun. Surv. Tut.}, vol.~13, no.~3, pp. 443--461,
  Third Quarter 2011.

\bibitem{zhu}
T.~Zhu, A.~Mohaisen, P.~Yi, and J.~Ma, \emph{Green Ad Hoc and sensor
  networks}.\hskip 1em plus 0.5em minus 0.4em\relax CRC Press, 2012, ch.~12,
  pp. 305--320.

\bibitem{raghun}
V.~Raghunathan, S.~Ganeriwal, and M.~Srivastava, ``Emerging techniques for long
  lived wireless sensor networks,'' \emph{IEEE Commun. Mag.}, vol.~44, no.~4,
  pp. 108--114, Apr. 2006.

\bibitem{kansal}
A.~Kansal, J.~Hsu, S.~Zahedi, and M.~B. Srivastava, ``Power management in
  energy harvesting sensor networks,'' \emph{ACM. Trans. Embed. Comput. Syst.},
  vol.~6, no.~4, pp. 1--8, Sep. 2007.

\bibitem{sharma}
V.~Sharma, U.~Mukherji, V.~Joseph, and S.~Gupta, ``Optimal energy management
  policies for energy harvesting sensor nodes,'' \emph{IEEE Trans. Wireless
  Commun.}, vol.~9, no.~4, pp. 1326--1336, Apr. 2010.

\bibitem{rajesh}
R.~Rajesh, V.~Sharma, and P.~Viswanath, ``Information capacity of energy
  harvesting sensor nodes,'' in \emph{Proc. IEEE Int. Symp. Inf. Theory
  (ISIT)}, Jul./Aug. 2011, pp. 2363--2367.

\bibitem{yang}
J.~Yang and S.~Ulukus, ``Optimal packet scheduling in an energy harvesting
  communication system,'' \emph{IEEE Trans. Commun.}, vol.~60, no.~1, pp.
  220--230, Jan. 2012.

\bibitem{yang_jcn}
------, ``Optimal packet scheduling in a multiple access channel with energy
  harvesting transmitters,'' \emph{Journal of Commun. and Net.}, vol.~14,
  no.~2, pp. 140--150, Apr. 2012.

\bibitem{yang_ozel}
J.~Yang, O.~Ozel, and S.~Ulukus, ``Broadcasting with an energy harvesting
  rechargeable transmitter,'' \emph{IEEE Trans. Wireless Commun.}, vol.~11,
  no.~2, pp. 571--583, Feb. 2012.

\bibitem{antepli}
M.~A. Antepli, E.~Uysal-Biyikoglu, and H.~Erkal, ``Optimal packet scheduling on
  an energy harvesting broadcast link,'' \emph{IEEE J. Sel. Areas Commun.},
  vol.~29, no.~8, pp. 1712--1731, Sep. 2011.

\bibitem{tutun}
K.~Tutuncuoglu and A.~Yener, ``Sum-rate optimal power policies for energy
  harvesting transmitters in an interference channel,'' \emph{Journal of
  Commun. and Net. Special Issue on Energy Harvesting in Wireless Net.},
  vol.~14, no.~2, pp. 151--161, Apr. 2012.

\bibitem{gunduz}
D.~Gunduz and B.~Devillers, ``Two-hop communication with energy harvesting,''
  in \emph{Proc. IEEE Int. Workshop on Comput. Advances in Multi-Sensor
  Adaptive Process. (CAMSAP)}, Dec. 2011, pp. 201--204.

\bibitem{ozel_Jsac}
O.~Ozel, K.~Tutuncuoglu, J.~Yang, S.~Ulukus, and A.~Yener, ``Transmission with
  energy harvesting nodes in fading wireless channels: optimal policies,''
  \emph{IEEE J. Sel. Areas Commun.}, vol.~29, no.~8, pp. 1732--1743, Sep. 2011.

\bibitem{Luo2013}
S.~Luo, R.~Zhang, and T.~J. Lim, ``Optimal save-then-transmit protocol for
  energy harvesting wireless transmitters,'' \emph{IEEE Trans. Wireless
  Commun.}, vol.~12, no.~3, pp. 1196--1207, Mar. 2013.

\bibitem{wcnc2013}
S.~Yin, E.~Zhang, J.~Li, L.~Yin, and S.~Li, ``Throughput optimization for
  self-powered wireless communications with variable energy harvesting rate,''
  in \emph{Proc. IEEE WCNC}, Apr. 2013, pp. 830--835.

\bibitem{zhang-WIPT}
R.~Zhang and C.~K. Ho, ``{MIMO} broadcasting for simultaneous wireless
  information and power transfer,'' \emph{IEEE Trans. Wireless Commun.},
  vol.~12, no.~5, pp. 1989--2001, May 2013.

\bibitem{cisco}
\BIBentryALTinterwordspacing
Cisco visual networking index: Global mobile data traffic forecast update,
  2012-2017. [Online]. Available:
  \url{www.cisco.com/en/US/solutions/collateral/ns341/ns525/ns537/ns705/ns827/white_paper_c11-520862.html}
\BIBentrySTDinterwordspacing

\bibitem{Fehske}
A.~Fehske, G.~Fettweis, J.~Malmodin, and G.~Biczok, ``The global footprint of
  mobile communications: The ecological and economic perspective,'' \emph{IEEE
  Commun. Mag.,}, vol.~49, no.~8, pp. 55--62, August 2011.

\bibitem{AMC}
A.~Duel-Hallen, ``Fading channel prediction for mobile radio adaptive
  transmission systems,'' \emph{Proc. IEEE}, vol.~95, no.~12, pp. 2299--2313,
  Dec 2007.

\bibitem{feedback}
A.~Ekpenyong and Y.-F. Huang, ``Feedback constraints for adaptive
  transmission,'' \emph{IEEE Signal Process. Mag.,}, vol.~24, no.~3, pp.
  69--78, May 2007.

\bibitem{Seong-selected}
K.~Seong, R.~Narasimhan, and J.~M. Cioffi, ``Queue proportional scheduling via
  geometric programming in fading broadcast channels,'' \emph{IEEE J. Sel.
  Areas Commun.}, vol. 24, no. 8, pp. 1593--1602, Aug. 2006.

\bibitem{Seong-isit}
K.~Seong, M.~Mohseni, and J.~M. Cioffi, ``Optimal resource allocation for
  {OFDMA} downlink system,'' in \emph{Proc. IEEE ISIT}, Jul. 2006, pp.
  1394--1398.

\bibitem{Jamali2014}
V.~Jamali, N.~Zlatanov, and R.~Schober, ``Adaptive mode selection for
  bidirectional relay networks - fixed rate transmission,'' \emph{Accepted for
  presentation at the IEEE Int. Conf. Commun. (ICC), Sydney, Australia}, Jun.
  2014.

\bibitem{ozel_TIT}
O.~Ozel and S.~Ulukus, ``Achieving {AWGN} capacity under stochastic energy
  harvesting,'' \emph{IEEE Trans Inf. Theory}, vol.~58, no.~10, pp. 6471--6483,
  Oct. 2012.

\bibitem{super-capasitor}
M.~Jayalakshmi and K.~Balasubramanian, ``Simple capacitors to
  supercapacitors\textemdash an overview,'' \emph{Int. J. Electrochem. Sci.},
  vol.~3, pp. 1196--1217, 2008.

\bibitem{Hoeffding}
W.~Hoeffding, ``Probability inequalities for sums of bounded random
  variables,'' \emph{Journal of the American Statistical Association}, vol.~58,
  no. 301, pp. 13--30, Mar. 1963.

\end{thebibliography}
\end{document}